\documentclass[twoside]{article}
\usepackage{fleqn,espcrc2}

\usepackage{graphicx}
\usepackage[figuresright]{rotating}

\newcommand{\AmS}{{\protect\the\textfont2
  A\kern-.1667em\lower.5ex\hbox{M}\kern-.125emS}}

\begin{document}

\title{Experimental fitting to the bipolaronic model of the
normal-state resistance of Bi$_2$Sr$_2$CaCu$_2$O$_8$ single
crystals%
\thanks{We would like to thank A. S. Alexandrov for a critical reading of the manuscript.
This research was supported by grants from the
National Sciences and Engineering Research Council (NSERC) of
Canada.}}

\author{Weimin Chen, J. P. Franck, and J. Jung %
\\Department of Physics, University of Alberta, Edmonton, Canada
T6G 2J1}


\begin{abstract}
Normal-state resistance data from Bi$_2$Sr$_2$CaCu$_2$O$_8$ single
crystals were fitted to the (bi)polaronic conduction model, $R=R_0
(T+\sigma_b T^2)/(1+bT)$, with satisfactory agreement over a wide
temperature range. The fluctuating conduction region is found to
be much narrower than that in the usual sense, as is the case for
a charged Bose-gas. We estimate the effective (bi)polaron mass to
be $\sim 4m_e$.
\vspace{1pc}
\end{abstract}

\maketitle

We report on the normal-state resistance measurement of
Bi$_2$Sr$_2$CaCu$_2$O$_8$ (Bi-2212) single crystals and its
fitting to the bipolaron model \cite{alexandrov,alexandrov2}. We
found that the data fitted to this theory over wide temperature
ranges. In some samples, the fitting spans from about $T_c + 40$ K
up to 300 K. Moreover, since the fitting range extends down to low
temperatures near $T_c$, the superconducting fluctuation region is
much narrower than that in the usual sense. These results provide
alternative interpretations of the peculiar behavior of the
normal-state resistivity.

The samples studied in the present work were Bi-2212 single
crystalline whiskers \cite{franck}, with typical dimensions of 2
to 3 mm $\times$ 10 $\times$ 0.5 $\mu$m along the $a, b,$ and $c$
axes, respectively. Resistance was measured by the standard
four-wire method in the $ab$-plane with a dc current of 0.5
$\mu$A.

Figure 1 shows a typical plot of resistance data. The dotted line
is a linear-$T$ fit at high temperature. As usual, the resistance
shows a downward deviation around 200 K, which is generally
associated with the pseudogap phenomenon (or superconducting
fluctuation). In the bipolaronic scenario, on the other hand, the
normal-state dynamics is dominated by small polarons and the
superconducting state is a charged (2$e$) Bose-Einstein condensate
of bipolarons \cite{alexandrov}. Quantitatively, this model gives
for the resistance:
\begin{equation}
R=R_0{T+\sigma_b T^2 \over 1+bT}, \label{R}
\end{equation}
where $\sigma_b$ is the relative boson-boson scattering cross
section, and $b$ is related to Hall coefficient,
\begin{equation}
R_H={v_0 \over 2e(n-n_L)(1+bT)}, \label{RH}
\end{equation}
with $(n-n_L)(1+bT)$ being the number of delocalized carriers in
the unit-cell volume $v_0$ (5.40 $\times$ 5.41 $\times$ 30.9
 \AA$^3$), and $R_0$ is a fitting parameter.

\begin{figure}[t]
\vspace{1pc}
\includegraphics[scale = 0.38]{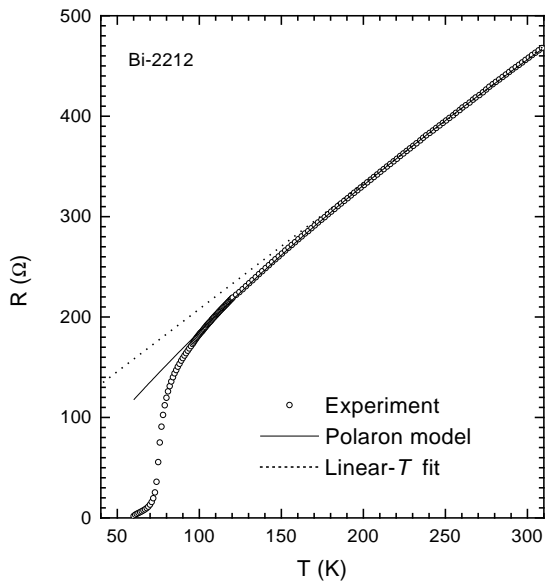}
\vspace{-2pc} \caption{Linear-$T$ (dotted line) and bipolaronic
(solid line) fittings to the normal-state resistance of a Bi-2212
sample. The latter covers a much wider region than the
former.\label{fig1}} \vspace{-0.8pc}
\end{figure}

By using the Hall coefficient data from Ref. \cite{hall}:
$R_H(10^{-3}$ cm$^3$/C) = 2.5, 2.3, 2.1, and 1.9, for $T$ = 150,
200, 250, and 300 K, we obtained the values for $b$ = 0.003
K$^{-1}$ and $(n - n_L) = 0.076$ from Eq.~\ref{RH}. This results
in $\sigma_b \approx$ 0.0011 to 0.0016 from initial fitting of six
samples. Suppose $\sigma_b$ is less sensitive to oxygen doping
than $b$, so we adopt constant $\sigma_b = 0.0015$ in the fitting.
The result is as shown by the solid line in Fig.~\ref{fig1}. As
can be seen, the data closely agrees with this model over a wide
temperature range. Two features in Fig.~\ref{fig1} are noticeable:
(1) While resistance deviation from the linear-$T$ fitting is
evident, it is well accounted for in the polaron model; and (2)
Superconducting fluctuation is seen only very close to $T_c$ in
the polaron fitting, in contrast to the much higher onset
temperature if the linear-$T$ criterion is used. The above
analysis, together with recent experiments which showed very
narrow coherent region above $T_c$ \cite{nature}, does not seem to
support superconducting fluctuation at high temperature ($\sim
200$ K). The coherence length obtained by using the
Aslamazov-Larkin model \cite{larkin} agrees well with literature
data. Therefore, we argue the superconducting fluctuation region
in the present picture is much narrower than that in the usual
sense. This point agrees with the scenario of a charged Bose-gas
\cite{alexandrov3}. A similar conclusion was also established from
transport measurement on YBCO \cite{narrowFluc}.

\begin{figure}[t]
\vspace{1.3pc}
\includegraphics[scale = 0.35]{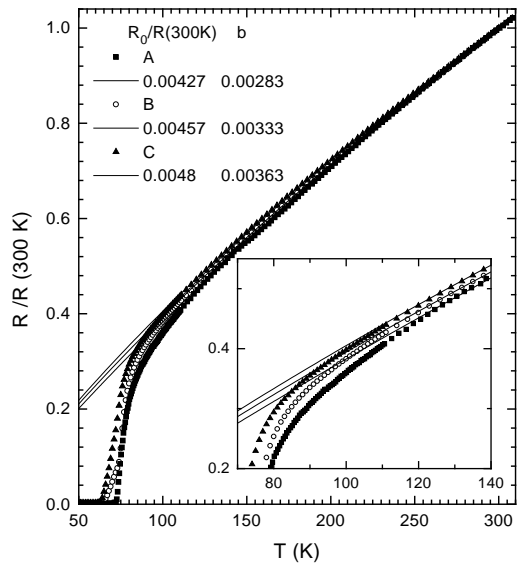}
\vspace{-1.8pc} \caption{Resistance data for three Bi-2212
samples. The numbers are fitting parameters of $R_0/R$(300 K) and
$b$ (refer to Eq.~\ref{R}). Inset: details near $T_c$.
\label{fig2}} \vspace{-0.22pc}
\end{figure}

The fitting results for three samples were shown in Fig.
\ref{fig2} (solid lines), all yielded mutual agreeable $b$ values.
The paraconducting deviation happens at $\sim(T_c + 50)$ K, which
is otherwise much higher ($\sim 200$ K) as described. The numbers
in the plot show the values of the fitting parameters
$R_0$/$R$(300 K) and $b$, respectively.

To further check the validity of this approach, we calculate the
effective mass of the in-plane boson mass \cite{alexandrov}, $m =
\pi \hbar^2/wa^2$, where the lattice constant $a =5.4$ \AA, $w$ is
related to $T_c$ through
\\
\begin{equation}
T_c={2w(n-n_L) \over \log\gamma^2}. \label{Tc}
\end{equation}
Take $T_c = 75$ K, $\gamma = 183$, and $(n - n_L) = 0.076$, we
have $m =4.27 \times$ free electron mass. Thus, the result may
well correspond to small polarons and to inter-site bipolarons.

\end{document}